\documentclass[aps,prl,preprint,superscriptaddress]{revtex4-1}
\usepackage{natbib}
\usepackage{amsmath}
\usepackage{graphicx}
\usepackage{nameref}

\usepackage{bm}%
\usepackage[colorlinks=true,linkcolor=blue]{hyperref}%

\usepackage{ulem}
\usepackage{setspace}
\def\url#1{\textcolor{blue}{\protect\small\sf #1}}
\usepackage{color}

\begin{document}

\title{On the Interpretation of Stratonovich Calculus}

%% ------------------------------------------------------------------------ %%
%
%  AUTHORS AND AFFILIATIONS
%
%% ------------------------------------------------------------------------ %%

% Method 1 (for all journals, except Reviews of Geophysics, which
% should use method 3):
% For three or fewer author/affiliation blocks, use \author{} and \affil{}

\author{W. Moon}
\email[]{wm275@damtp.cam.ac.uk}
\affiliation{Institute of Theoretical Geophysics, Department of Applied Mathematics  \& Theoretical Physics, University of Cambridge, Cambridge CB3 0WA, UK}
\affiliation{Yale University, New Haven, CT, 06520, USA}

\author{J. S. Wettlaufer}
\email[]{wettlaufer@maths.ox.ac.uk}
\affiliation{Mathematical Institute, University of Oxford, Oxford OX2 6GG, UK}
\affiliation{Yale University, New Haven, CT, 06520, USA}
%\affiliation{Nordic Institute for Theoretical Physics (NORDITA), 10691 Stockholm, Sweden}

%\date{today}

\begin{abstract}

The It\^{o}-Stratonovich dilemma is revisited from the perspective of the interpretation of Stratonovich calculus using shot noise.   
Over the long time scales of the displacement of an observable, the principal issue is how to deal with finite/zero autocorrelation of the stochastic noise.  The former (non-zero) noise autocorrelation structure preserves the normal chain rule using a mid-point selection scheme, which is the basis Stratonovich calculus, whereas the instantaneous autocorrelation structure of It\^{o}'s approach does not.  By considering the finite decay of the noise correlations on time scales very short relative to the overall displacement times of the observable, we suggest a generalization of the integral Taylor expansion criterion of Wong and Zakai \cite{wong1965} for the validity of the Stratonovich approach.

\end{abstract}

%% ------------------------------------------------------------------------ %%
%
%  TEXT
%
%% ------------------------------------------------------------------------ %%

%\begin{article}

\maketitle 

\section{Introduction}

Stochastic dynamical models are basic to the understanding of the role of random forcing in a wide range of scientific and engineering systems  \cite[e.g.,][]{van1976, hasselmann1976, benzi1981}.  The central theoretical approaches arise from the Einstein and Langevin studies of Brownian motion \cite{einstein1905, langevin1908}, which provide mathematically different but physically equivalent and complimentary descriptions of the fate of a body (a pollen particle in water observed under a microscope) under the influence of random non-deterministic collisions (water molecules).   Einstein determined the time evolution of the probability density of particles by solving the Fokker-Planck equation whereas Langevin wrote down an explicit deterministic momentum equation for a particle augmented by a Gaussian white noise forcing which perturbs the particle trajectory.  The approach of Langevin now constitutes a 
 canonical  ``stochastic differential equation'' (SDE) for a daunting scope of systems, but since its introduction the mathematical and physical interpretation of the noise term has been discussed and debated. 

Whether viewing the problem from configuration space through the solution of the Fokker-Planck equation, solving a Langevin equation, or generating statistical realizations by performing Monte-Carlo simulations of the particles under consideration \cite[e.g.,][]{kloeden1977}, a consistent interpretation and calculational scheme of the noise structure is needed.  The approach developed by It\^{o} \cite{ito1944} rests upon the Markovian and Martingale properties; the former captures the concept of a ``memoryless'' process, wherein the conditional probability distribution of future states depends solely on the present state and the latter that, given all prior events, the expectation value of future stochastic events equals the present value.  These properties have the advantage of  simplifying many complicated time integrals but the disadvantage of requiring a new calculus which does not obey the traditional chain rule.  In contrast, the approach of Stratonovich\cite{stratonovich1966} does not invoke the Martingale property and preserves the chain rule and allows white noise to be treated as a regular derivative of a Brownian (or Weiner) process, $W_t$. 

A model SDE (to which we shall return later) for a variable $x(t)$ is 
\begin{eqnarray} 
\label{eqn:SDE}
  \frac{dx}{dt} = a(x,t)+b(x,t)\frac{dW_t}{dt},
 \end{eqnarray}
wherein the first term is ``deterministic'' and the second term as ``stochastic''.  The case in which $b(x,t)$ is constant is referred to as additive noise and when it depends on $x=x(t)$ it is multiplicative noise.  Now, despite being able to integrate equation (\ref{eqn:SDE}) in a formal sense as 
\begin{equation} 
\label{eqn:SDE_int}
x(t) - x(t_0) = \int_{t_0}^{t}  a(x,t^\prime)  dt^\prime  + \int_{t_0}^{t} b(x, t^\prime) dW_t (t^\prime), 
 \end{equation}
the crux of the ``It\^{o}-Stratonovich dilemma'' resides in the integral over the Brownian process $W_t$. The issue is laid bare by recalling the definition of the definite integral of a real valued function $f(t)$ in terms of the Riemann sum viz., 
\begin{equation} 
\label{eqn:integraldefn}
\int_{t_0}^{t}  f(t^\prime)  dt^\prime \equiv \lim_{n \rightarrow \infty} \sum_{j=0}^{n-1} f(\tilde{t}_k) (t_{k+1} - t_k), 
\end{equation}
where importantly $\tilde{t}_k \in [t_{k}, t_{k+1}]$.  The issues are ({\em i}) the Brownian process $W_t$ is nowhere differentiable {\em and} because $\frac{dW_t}{dt}$ is $\delta$-autocorrelated then in any interval on the real line the white noise it represents fluctuates an infinite number of times with an infinite variance,  ({\em ii}) the defining limit of the integral in equation (\ref{eqn:integraldefn}) depends on the place in $[t_{k}, t_{k+1}]$ where $\tilde{t}_k$ is chosen, the choice that provides the seed of the dilemma and the origin of the two different calculi.  

In the It\^{o} approach the choice is $\tilde{t}_k = t_{k}$, which maintains the Martingale property due to the fact that $t_k$ is the present value in the integrand, thereby forcing  the expectation value in $[t_{k}, t_{k+1}]$ to the present value.  In contrast, the Stratonovich approach defines $\tilde{t}_k = (t_{k} + t_{k+1})/2$, which abandons the Martingale property but maintains the normal rules of calculus.   Stratonovich referred to his choice as a ``symmetrization'' between past and future.  Despite its wide usage, we have not found a physical interpretation of the basis of the past/future symmetrization of Stratonovich calculus.  

The outline of this note is as follows.  In the next section we summarize our approach and make brief mention of its connection to the dilemma generally and other studies.  We then outline the conventional viewpoint and the perspective of stochastic calculus, before coming to our main point and then concluding.  

\section{A Brief Comparison and Contrast}

Wong and Zakai \cite{wong1965} argued that in any real world system perfect white noise does not exist and that Brownian motion $x(t)$ approximates a description $x_n(t)$ that is continuous with at least a piece-wise continuous derivative.  By showing that $x_n(t) \rightarrow x(t)$ as $n \rightarrow \infty$ they recovered Stratonovich calculus.  Accordingly, the choice of stochastic calculus resides in the characteristics of the noise and continuity arguments.  In finance, it is argued that short time scale processes are truly discontinuous and thus  It\^{o} calculus is preferred \cite[e.g.,][]{shreve2004}, thereby maintaining the Martingale property.  However, in physics the continuous motion of Brownian particles influenced by high frequency white noise has long been considered within the framework of normal calculus \cite{uhlenbeck1930}.  Conceptually, there is no clear distinction between the statistics of  water molecules colliding pollen grains and trading options or stocks.  Hence, the question remains if, how and when it is appropriate to use continuity considerations as a core criterion to choose either of the calculi being discussed here. 

As noted in the introduction, the complimentary approaches of Einstein and Langevin, with a reliance on  $\delta$-autocorrelated noise, provide consistent testable predictions of Brownian motion.  The combination of this and the  additional consistency with Stratonovich calculus it has been suggested that physical scientists should avoid It\^{o} calculus  \cite{west1979}.  Although Van Kampen \cite{van1981} cautiously suggested that the Langevin equation is  intrinsically insufficient for representing systems with internal noise, it is still the case that Stratonovich calculus is appropriate for both internal and external noise and there are an enormity of problems in which one cannot make this distinction.  

Here we focus on an ostensibly physical argument to discuss the origin of Stratonovich calculus. We generalize the theorem Wong and Zakai \cite{wong1965} using the integral Taylor expansion and appealing solely to the $L^2$ integrability of a function, thereby avoiding discussions of the regularity of a stochastic process.  Because the principal difference between the It\^{o}-Langevin and the Stratonovich-Langevin formulations lies in the drift term, we focus on this in our approach and examine the intrinsic nature of short time scale processes approximated by white noise to decide which calculus is more appropriate. As opposed to Turelli \cite{turelli1977}, who took as a continuous deterministic model an approximation of a discretized model (in population dynamics), we begin with a well defined deterministic model.  
  
\section{The Conventional Physical Science Perspective}

The term ``white noise'' refers to a random signal $\Gamma(t)$ in which all frequencies contribute equally to the power spectral density.  For a general stationary random process the autocorrelation ${\cal R}(t^\prime - t) = \langle\Gamma(t')\Gamma(t)\rangle$, where $\langle\cdot\rangle$ is the ensemble average, is definition independent of time and symmetric/even ${\cal R}(t^\prime - t) = {\cal R}(t - t^\prime)$.   If $t - t^\prime \equiv \tau$, then let  $\tau_c$ be the de-correlation time such that ${\cal R}(\tau_c) = \mbox{e}^{-1} {\cal R}(0)$.  Typically, as $\tau_c \rightarrow 0$ but $\int_{- \infty}^{\infty} {\cal R}(\tau_c) d\tau_c$ remains finite, we say that ${\cal R}(\tau_c) = \delta(0)$ and the white noise has a $\delta$-autocorrelation structure. 

In science and engineering the traditional form of the Langevin equation, slightly different than that in equation (\ref{eqn:SDE}),  is written as
 \begin{eqnarray} \label{eqn:sto_eqn}
  \frac{dx}{dt} = a(x,t)+b(x,t)\Gamma(t),
 \end{eqnarray}
where the noise forcing is $\delta$-autocorrelated as $\langle\Gamma(t')\Gamma(t)\rangle = \delta(t'-t)$.  Based on this, we calculate $\langle\frac{dx}{dt}\rangle$ starting with the first integral 
\begin{eqnarray} \label{eqn:sto_eqn_white}
 x(t+\tau)-x(t) = \int_{t}^{t+\tau}\left(a[x(t'),t']+b[x(t'),t']\Gamma(t')\right)dt'.
\end{eqnarray}
We assume that $\tau$ is small relative to the time over which the macroscopic system evolves, but  large relative to the time scale of fluctuations in the stochastic process  
$\Gamma(t)$. When $a(x,t)$ and $b(x,t)$ are analytic, we can Taylor expand them about a time $t$  as 
\begin{eqnarray} 
\label{eqn:ab}
 &a(x(t'),t')=a(x(t),t)+\frac{\partial a}{\partial x}(x(t')-x(t))+\frac{\partial a}{\partial t}(t'-t)+\ldots \nonumber \\
 &b(x(t'),t')=b(x(t),t)+\frac{\partial b}{\partial x}(x(t')-x(t))+\frac{\partial b}{\partial t}(t'-t)+\ldots .
\end{eqnarray}
following Risken \cite{risken1996}.  
Inserting (\ref{eqn:ab}) into (\ref{eqn:sto_eqn_white}) leads to
\begin{eqnarray} \label{eqn:taylor_white}
 x(t+\tau)-x(t) &= a(x(t),t)\tau+\frac{\partial a}{\partial x}\int_{t}^{t+\tau}(x(t')-x(t))\,dt'+\frac{\partial a}{\partial t}\frac{1}{2}\tau^2 \nonumber \\
   &+b(x(t),t)\int_{t}^{t+\tau}\Gamma(t')dt'+\frac{\partial b}{\partial x}\int_{t}^{t+\tau}(x(t')-x(t))\Gamma(t')\,dt' \nonumber \\
   &+\frac{\partial b}{\partial t}\int_{t}^{t+\tau}(t'-t)\Gamma(t')\,dt'+ h.o.t 
\end{eqnarray}
For $x(t')-x(t)$, we can treat equation (\ref{eqn:sto_eqn_white}) iteratively and then take the ensemble average, which is
\begin{eqnarray} \label{eqn:taylor_final_white}
 \langle x(t+\tau)-x(t)\rangle 
 = a(x,t)\tau + b\frac{\partial b}{\partial x}\int_{t}^{t+\tau}\,dt'\int_{t}^{t'}\langle\Gamma(t'')\Gamma(t')\rangle dt'' 
 + O(\tau^2).
\end{eqnarray}
Therefore,
\begin{eqnarray} \label{eqn:taylor_final_white_order}
 \lim_{\tau \to 0}\frac{1}{\tau}\langle x(t+\tau)-x(t)\rangle = a(x,t)+b\frac{\partial b}{\partial x}
 \frac{1}{\tau}\int_{t}^{t+\tau}\,dt'\int_{t}^{t'}\langle\Gamma(t'')\Gamma(t')\rangle dt''.
\end{eqnarray}
By the definition of white noise discussed above, $\langle\Gamma(t')\Gamma(t'')\rangle=\delta(t'-t'')$, we have
\begin{eqnarray}
 \int_{t}^{t+\tau}\,dt'\int_{t}^{t'}\langle \Gamma(t'')\Gamma(t')\rangle dt''=\int_{t}^{t+\tau}\,dt'\int_{t}^{t'}\delta(t''-t')dt''.
\end{eqnarray}
In the integral $\int_{t}^{t'}\delta(t''-t')dt''$, half of the contribution of the delta function is considered such that 
the value of the integral becomes $\frac{1}{2}$. Therefore, $\int_{t}^{t+\tau}\,dt'\int_{t}^{t'}\langle\Gamma(t'')\Gamma(t')\rangle dt''$
is equal to $\frac{1}{2}\tau$. What is the physical meaning of the half
contribution of the delta function? The entire interval $[t,t']$ does not contribute to the integral $\int_{t}^{t'}\delta(t''-t')dt''$. We shrink the interval near $t'$ as $[t'-\epsilon, t']$, where $\epsilon$ can be interpreted as the size of a sub-interval in the Riemann summation  discussed above in distinguishing the stochastic calculi. 
Because the delta function is even $\delta(t''-t')=\delta(t'-t'')$, the half contribution can be written as
\begin{eqnarray}
 \int_{t}^{t'}\delta(t''-t')dt'' = \frac{1}{2}\int_{t'-\epsilon}^{t'+\epsilon}\delta(t''-t')dt''.
\end{eqnarray}
Despite this being just a simple mathematical modification, we can interpret it as a representation that the weighting from past is equal to that from future around $t'$. 
The origin of this symmetry comes from the time 
symmetry of the autocorrelation function ${\cal R}(\tau)$. We will connect this half contribution of the autocorrelation to the Stratonovich integral below, but we note now the above development yields 
\begin{eqnarray}
\left<\frac{dx}{dt}\right> = a(x,t)+\frac{1}{2}b(x,t)\frac{\partial}{\partial x}b(x,t), 
\end{eqnarray}
which is exactly same as that obtained from Stratonovich calculus.

\section{The Stochastic Calculus Perspective}

Mathematicians have long questioned the validity of the conventional definition of white noise because Brownian motion 
$W_t$ is no where differentiable.  Namely, defining $\Gamma(t)$ as $\frac{dW_t}{dt}$ is poorly grounded in normal calculus and we must revisit the interpretation of equation (\ref{eqn:sto_eqn}) 
in light of the mathematical definition of Brownian motion $W_t$; 
\begin{enumerate}
 \item $W_0$ = 0
 \item $W_t$ is almost surely continuous (or sample-continuous) with respect to $t$
 \item $W_t$ has independent increments and $W_t-W_s$, with $s<t$, is governed by a normal distribution, $N(0,t-s)$, with zero mean and standard deviation $\sqrt{t-s}$.
\end{enumerate}
Therefore, equation (\ref{eqn:sto_eqn}) can be rewritten as
\begin{eqnarray}
 dx = a(x,t)dt+b(x,t)dW_t.
\end{eqnarray}
On the one hand, using this integral form avoids the issue of the differentiability of  $W_t$, but this comes at the cost of having to introduce a different sort of calculus.
However, before coming to the issue of the implementation of a particular stochastic calculus, we perform the same expansion as discussed previously, in (\ref{eqn:taylor_white}) and (\ref{eqn:taylor_final_white}), which leads to
\begin{eqnarray}
 \label{eqn:meansquare}
 \langle x(t+\tau)-x(t) \rangle 
 = a(x,t)\tau+b(x,t)\frac{\partial}{\partial x}b(x,t)\left<\int_{t}^{t+\tau}\int_{t}^{t'}dW''_t dW'_t \right>+ O(\tau^2).
\end{eqnarray}
The first term $a(x,t)$ can be interpreted as the deterministic response to macro-scale forcing and the second term embodies the interaction between a macroscopic
quantity $b(x,t)$ and the microscopic accumulation of random noise forcing. On a time $\tau$ that is short from the macroscopic perspective, 
the interaction between two scales manifests itself multiplicatively. The contribution of the micro-scale to the change of
the macroscopic quantity $x$ appears as a double integral of the Brownian motion. It is at this point that we must decide upon a specific
calculus, and this choice determines the value of the double
integral in equation (\ref{eqn:meansquare}).
%Hence, we need a proper interpretation of $\int_{t}^{t+\tau}\int_{t}^{t'}dW''dW'$ with the microscopic eye to find a appropriate value of the integral. 

The double integral can be written as $\int_{t}^{t+\tau}W'dW'$ and approximated by a Riemann sum as 
\begin{eqnarray}
 \int_{t}^{t+\tau}W'dW'=\sum_{k=0}^{N-1}W_{t'_k}\Delta W_{t_k},
\end{eqnarray}
where $\Delta W_{t_k} \equiv W_{t_{k+1}}-W_{t_{k}}$. As should be intuitive from the previous discussion, the confusion lies in the choice of $t'_k$ for $W_{t'_k}$. As can be seen below, this choice determines the average of the integral viz., 
\begin{eqnarray}
 \left<\sum_{k=0}^{k=N-1}W_{t'_k}\Delta W_{t_k}\right>&=\sum_{k=0}^{k=N-1}\left<(W_{t'_k}-W_{t_k}+W_{t_k})(W_{t_{k+1}}-W_{t_k})\right> \nonumber \\
  &=\sum_{k=0}^{k=N-1}\left<(W_{t'_k}-W_{t_k})(W_{t_{k+1}}-W_{t_k})\right> \nonumber \\
  &=\sum_{k=0}^{k=N-1}(t'_k-t_k).
\end{eqnarray}
It\^{o} \cite{ito1944} argued that by choosing $t'_k$ = $t_k$  the most important characteristic preserved is the Martingale property.  This implies that
the expectation value of a future event is equivalent to that of the present value in the sub-interval $[t_k, t_{k+1}]$. 
Hence, the expectation value of the double integral should be zero. The disadvantage of It\^{o} calculus is that the chain rule of  normal calculus is lost; one needs additional terms to preserve the Martingale property during integration. On the other hand, Stratonovich \cite{stratonovich1966} chose $t'_k$ = $\frac{t_k+t_{k+1}}{2}$, 
and the integral has a value of $\frac{1}{2}\tau$. This approach preserves the chain rule, implying that Stratonovich calculus must be used when 
we approximate noise processes that are continuous and have piece-wise continuous derivatives \cite[e.g.,][]{wong1965}.  When we approximate the rapid fluctuations on a stochastic process as white noise, then we must preserve the normal rules of calculus if we are to capture the dynamics of Brownian motion. 

In a classical physical world where the continuity is more or less guaranteed, Stratonovich calculus is evidently the most appropriate. However, when we consider the colliding of particles or the buying or selling of stocks or options, these short time-scale phenomena {\em appear} to be discontinuous.  Such circumstances lead us to question the validity of the continuity of a stochastic process. That said, it may be impossible to find events that have zero autocorrelation.  Thus it is a natural question to examine the effect of a non-zero autocorrelation in a multiplicative noise process by investigating limiting behaviors, which may provide some insight into the nature of the appropriate stochastic calculus.  Hence we consider the simplest case of multiplicitive behavior; $dx/dt=xF(t)$, where $F(t)$ is 
a stochastic process with a finite decorrelation time. For this example, using the same procedure used to arrive at equation (\ref{eqn:taylor_final_white}), we find that 
\begin{eqnarray}
 \langle x(t+\frac{\tau}{n})-x(t)\rangle =x(t)\int_{t}^{t+\frac{\tau}{n}}\int_{t}^{t'}\langle F(t'')F(t')\rangle dt''dt'.
\label{eqn:idealmult1}
\end{eqnarray}
An important feature of our argument involves the time $\tau/n$, with integer $n$, which is a subdivision of the total domain $[t,t+\tau]$. 
For example, consider the commonly used autocorrelation $\langle F(t'')F(t')\rangle=\frac{\sigma^2}{2\tau_c}\mbox{e}^{-|t''-t'|/\tau_c}$,
where $\tau_c$ is the decorrelation time of the amplitude of the stochastic noise $F(t)$ satisfying $\tau_c <<  \tau$.  Now, let $\tau_{\mbox{m}}\equiv\tau/n$ be a minimum time over which we can observe the decay of the stochastic noise. 

Whereas it is always the case that $\tau >> \tau_c$, the value of $\tau_{\mbox{m}}$ depends on the fineness $n$ of the subdivision, with $\tau_{\mbox{m}}$ {\em increasing} as $n$ {\em decreases}; coarsening the temporal resolution. So long as $\tau_{\mbox{m}} >> \tau_c$ we can still justify the use of the stochastic differential equation but once $\tau_{\mbox{m}} \leq \tau_c$ this is no longer the case.  Therefore, increasing $\tau_{\mbox{m}}$ to values above $\tau_c$ but still small relative to $\tau$ one reaches a value of $\tau_{\mbox{m}}\equiv\tau_{\star}$ that is sufficiently short that the decay of the noise can be observed but sufficiently large so that the stochastic differential equation is valid.  This is achieved as follows.  Consider $n$ such that $\tau >> \tau_{\mbox{m}} = \tau_c$, which we rewrite as $1 >> \frac{1}{n} = \frac{\tau_c}{\tau}$.  
Now, as $n$ decreases from $\frac{\tau}{\tau_c}$ it will reach a value $n=n_{\star}$ such that $1 >> \frac{1}{n_{\star}} >> \frac{\tau_c}{\tau}$ and our analysis is valid for $n_{\star} \ge n >> 1$ as described presently.  

With the above form of the autocorrelation equation (\ref{eqn:idealmult1}) becomes
\begin{eqnarray}
 \langle x(t+\frac{\tau}{n})-x(t)\rangle
 =x(t)\times\frac{1}{2}\sigma^2\frac{\tau}{n}\left[1+\frac{\tau_c}{\tau/n}\left(e^{-\frac{\tau}{n\tau_c}}-1\right)\right] ,
%\label{eqn:idealmult2}
\label{eqn:multiplicity}
\end{eqnarray}
which for $n_{\star} \ge n >> 1$ can be approximated 
as $\langle x(t+\frac{\tau}{n})\rangle =x(t)(1+\frac{1}{2}\sigma^2\frac{\tau}{n})$ and then rewritten as
\begin{eqnarray}
 \langle x(t+\tau) \rangle = x(t)\left(1+\frac{1}{2}\sigma^2\frac{\tau}{n}\right)^n.
\end{eqnarray}
The limit $n_{\star} \ge n >> 1$ insures that the right hand side converges to $x(t)e^{\frac{1}{2}\sigma^2\tau}$ and hence 
\begin{eqnarray}
 \langle x(t+\tau) \rangle=x(t) e^{\frac{1}{2}\sigma^2\tau} \approx x(t)(1+\frac{1}{2}\sigma^2\tau),
\label{eqn:strat}
\end{eqnarray}
consistent with the result from Stratonovich calculus, in which $\frac{1}{2}\sigma^2{\tau}$ is analogous 
to $\langle W_t \Delta W_t \rangle$ in the Riemann sum with $W_t=W_{(t_k+t_{k+1})/2}$ and $\Delta W_t=W_{t_{k+1}}-W_{t_k}$. This is the essence of 
Stratonovich calculus; choosing the mid-point in a subinterval is equivalent to including only {\em half of the effect of a finite 
autocorrelation}. 
Clearly, in this example, the magnitude of $\tau_c$ does not play a role unless $\tau_{\mbox{m}} \searrow \tau_c$ and the neccessary separation of time scales starts to be violated. Therefore, in this case
the commonly-used definition of the white noise $\langle F(t'')F(t')\rangle\propto \delta (t''-t')$ becomes an increasingly better approximation, which we can understand as follows. 

Recall that
\begin{eqnarray}
 \int_{t}^{t'}\frac{\sigma^2}{2\tau_c}e^{-\frac{|t''-t'|}{\tau_c}}dt'' \approx \int_{t}^{t'}\sigma^2\delta(t''-t')dt''\sim {\cal O}(\sigma^2).
\end{eqnarray}
We can proceed along the same lines as above but without the need to invoke a specific form of the autocorrelation.  Rather, we only need to assume that $F(t)$ is 
stationary, 
\begin{eqnarray}
 \int_{t}^{t+\frac{\tau}{n}}\int_{t}^{t'}\langle F(t'')F(t')\rangle dt''dt'
 &= \int_{-\tau/n}^{0}ds\int_{t-s}^{t+\tau/n}K(s) dt'=\int_{-\tau/n}^{0}K(s)(\tau+s) ds\nonumber \\
  & \approx \frac{1}{2}\frac{\tau}{n}\int_{-\infty}^{\infty}K(s)ds,
\end{eqnarray}
where $K(s)=\langle F(t)F(t+s) \rangle$ and we exploited $\tau/n >> \tau_c$, thereby neglecting $s$ compared to $\tau/n$ (see e.g., Ch. 15 of \cite{Reif}). 
Here,
$\frac{1}{2}\frac{\tau}{n}\int_{-\infty}^{\infty}K(s)ds$ is an approximation of $\langle W_t\Delta W_t\rangle$ in the Riemann sum, where the appropriate form of stochastic calculus is determined by the magnitude of $\int_{-\infty}^{\infty}K(s)ds$. Namely, if 
\begin{eqnarray}
 \int_{-\infty}^{\infty}K(s)ds \sim {\cal O}(\sigma^2),
\end{eqnarray}
where $\langle F^2(t)\rangle=\sigma^2$, the determination of $\langle W_t\Delta W_t\rangle$ 
follows Stratonovich calculus and hence choosing a $\delta$-function for $K(s)$ is 
appropriate. However, if 
\begin{eqnarray}
 \int_{-\infty}^{\infty}K(s)ds << \sigma^2,
\label{eqn:itotime}
\end{eqnarray}
then $\frac{1}{2}\frac{\tau}{n}\int_{-\infty}^{\infty}K(s)ds$ is negligible and the determination of $\langle W_t\Delta W_t\rangle$ follows
 It\^{o}-calculus.  Here, $K(s)$ can be considered approximately as a zero-measure function in $L^2$ space such that the $\delta$-function cannot be used 
for the approximation of the autocorrelation function $K(s)$. This argument is similar to Morita's \cite{morita1981}, who also considered three 
time scales and his choice of stochastic calculus also relies on the condition for the autocorrelation function, but as manifests itself in the form of 
a generalized Fokker-Planck equation.  

The choice of Stratonovich calculus for colored noise has been considered within the framework of a master equation by Horsthemke and Lefever \cite{horsthemke2006}. 
Colored noise converges to Gaussian white noise through the application of the central limit theorem, which they argued through a rescaling of time ($\Delta t$) and space ($\Delta x$) assuming a diffusive process, viz., $\Delta t \rightarrow \Delta t/\epsilon^2, \Delta x \rightarrow \Delta x/\epsilon$. In the limit $\epsilon \rightarrow 0$ the master equation becomes the Fokker-Planck equation with Gaussian white noise interpreted using Stratonovich calculus.  Horsthemke and Lefever note that the interpretation of Stratonovich calculus in the approach of Wong and Zakai \cite{wong1965} requires the strong regularity of colored noise (a C$_1$ function) for convergence to Gaussian white noise.  However, Wong and Zakai \cite{wong1965} demonstrate the convergence directly from an SDE, whereas Horsthemke and Lefever \cite{horsthemke2006} avoid the issue of regularity by beginning with an Uhlenbeck-Ornstein process, which is not C$_1$, and demonstrate convergence using a master equation, but do not connect the finite autocorrelation of colored noise with a physical interpretation that underlies the construction of Stratonovich calculus.  

It is the mid-point selection rule in the Riemann sum that is associated with the finite noise autocorrelation.   However, traditionally this selection rule has been interpreted as the ``magical'' choice required in order to recover the normal rules of calculus, but Stratonovich calculus itself does not provide any new dynamical insight that plays a role analogous to the Martingale property of It\^{o} calculus.  Hence, we seek a basic understanding of how the mid-point selection rule emerges as a {\em consequence} of applying the central limit theorem to colored noise.   Our approach is to consider shot noise, which has an auto-correlation represented by a 
$\delta$-function.

What is the origin of the $\delta$-function?  Brownian motion can be generated by a collection of independent random processes.  In the Riemann-summation approximation of the integral $\int_{t}^{t+\tau}W'dW'$, we first considered the subdivision of the time domain $[t,t+\tau]$ using $\tau/n$ with integer $n$.   As discussed in the argument leading to equation (\ref{eqn:strat}), so long as it is large the freedom to choose $n$ determined the time scale $\tau/n$ over which the decay of the noise forcing is observed; when $\tau/n >> \tau_c$ then the noise signal appears as a discrete packet, appropriately simulated as shot noise \cite[e.g.,][]{lathi1998}. 

Consider two ``square'' signals with amplitude $h$ and time duration $w$, occurring with probability $\alpha w$.  When both are contained in the same packet and have the same sign ($\pm h$) they are positively correlated.  Otherwise they are independent and uncorrelated. In the probability domain $\Omega$ the autocorrelation ${\cal R}(\tau)$  is 
\begin{eqnarray}
 {\cal R}(\tau)&=\int_{\Omega} n_1n_2P(n_1,n_2)dn_1dn_2,
\end{eqnarray}
where $n_1$ ($n_2$) is a random variable occurring at the time $t$ ($t+\tau$) and $P(n_1,n_2)$ is the joint probability density function of $n_1$ and $n_2$. We need only consider the case when $\tau$ is smaller than $w$, which is
\begin{eqnarray}
 {\cal R}(\tau)&= \int_{\Omega} n^2_1 P(n_1)P(n_2|n_1)dn_1dn_2 \nonumber \\
                       &=\left(1-\frac{\tau}{w}\right)\alpha w \left(\frac{1}{2}\left(\frac{h}{w}\right)^2+\frac{1}{2}\left(-\frac{h}{w}\right)^2\right) \nonumber \\
                       &={\alpha}h^2\frac{1}{w}\left(1-\frac{\tau}{w}\right), 
\end{eqnarray}
from which we can readily calculate
\begin{eqnarray}
\int_{-\infty}^{\infty}R(\tau)d\tau\approx {\alpha}h^2\int_{-w}^{w}\frac{1}{w}\left(1-\frac{\tau}{w}\right)d\tau=2{\alpha}h^2.
\end{eqnarray}
Thus, the constant $\alpha h^2$ corresponds to $\sigma^2$, $h/w$ is analogous to $dW/dt$, and hence ${\cal R}(\tau) = 2\sigma^2\delta(\tau)$. 
Therefore, despite our argument emerging from shot noise, it leads naturally to the definition of  white noise in which there exists a finite time correlation that is too minute to be realized when viewed from a coarse time scale, a coarseness which depends on the observer. 

Clearly, a similar procedure applies directly to the Riemann-sum so long as the same constraints on the subdivision of the time domain are in place. 
When $\tau/n \simeq w$, two signals of the same sign are positively correlated in the interval. Hence, the 
random variables $n_1 n_2$ only have the value $h^2$ with probability $\int_{\Omega}P(n_1)P(n_2|n_1)dn_1 dn_2$, where $P(n_1)$ is $\alpha \tau/n$ and $P(n_2|n_1)$ is $1/2$.
Because the conditional probability $P(n_2|n_1)$ is equal to the fraction of $n_2>n_1$ in the domain $\{(n_1,n_2)| 0<n_1<\tau/n, 0<n_2<\tau/n\}$, $n_1 n_2$ take the value $h^2$ with probability $\frac{1}{2}\alpha \frac{\tau}{n}$ and the value $0$ with the probability $1-\frac{1}{2}\alpha \frac{\tau}{n}$. According to the central 
limit theorem, a very large collection of realizations $n_1 n_2$ is given by a Gaussian distribution with mean $\frac{1}{2}\alpha h^2 \tau$.  This result is analogous to $\frac{1}{2}\sigma^2 W^2_t$ which was obtained from the Stratonovich calculus integral of $\int_{0}^{t} W_t dW_t$.  Moreover, the factor of $1/2$ originates in the time order of $n_1$ and $n_2$ which is traced to the constraint $n_1 < n_2$ when $P(n_2|n_1)$ is calculated. In other words, the present value $n_1$ is influenced in a time-symmetric manner and this factor of $1/2$ reflects this symmetry through a properly ordered time integration.  

In this simple argument one sees that the $\delta$-autocorrelated noise (used in the definition of white noise) and the Stratonovich calculus approach have the same origin. 
Both formalisms use different approaches to capture the accumulated influence of  short time scale correlations of a noise source that are not represented in the long time-scale dynamics.   Importantly, temporal continuity of the noise is not a necessary condition. In order to demonstrate this, we used discontinuous auto-correlated shot noise to recover both the $\delta$-function definition of  white noise and the mid-point selection procedure of  Stratonovich calculus.
Moreover, one can make the simple case more realistic by treating $h$ as a random variable and the same logic holds. 

Finally, we note that despite our pedagogical example $dx/dt=xF(t)$, the results are general. Thus, when we replace $x$ in the right hand side with 
$b(x,t)\frac{\partial}{\partial x}b(x,t)$, we can generalize our argument to any multiplicative noise case. There are of course a myriad of ways to decide which calculus is most appropriate to the problem at hand.  Hence, whereas the integral Taylor expansion method enables one to only focus on the nature of the short time-scale processes,
the comparison between the integral of autocorrelation and the variance of the short time-scale process focuses the choice on the specific mathematical model or scientific problem.   Of particular interest to us is the question of how the stability of the non-autonomous SDE's of interest in climate dynamics are influenced by such considerations \cite{MW2013}.

\section{Related Approaches}

According to our result using shot noise, even an infinitesimal noise correlation in a stochastic differential equation can be interpreted  using  Stratonovich calculus. One might consider our analysis as a generalised version of the Wong and Zakai \cite{wong1965} approach because shot noise here is not a C$_1$ function. 
Considering the ubiquity of colored noise in real systems, 
It\^{o} calculus might be interpreted as an idealized mathematical procedure, only applicable to true white noise processes that are never realized in nature.
Hence, we ask whether there are situations in which It\^{o} calculus can be used for colored noise?

The conceptual model we are dealing with in stochastic differential equations is that we have a principle deterministic process whose fate is influenced by short time scale fluctuations, which are considered as noise.  In building a mathematical model it is common to ignore the influence of the short time scale processes on the deterministic dynamics, but there are situations when this may be known to be a poor assumption such as is the presence of inertial \cite{Kupferman:2004} or feedback \cite{Pesce:2013} effects.  Indeed, Kupferman and colleagues  \cite{Kupferman:2004} studied systems with multiplicative {\em colored} noise and inertia to find that if the correlation time of the noise is faster (slower) than the relaxation time, this leads to the It\^{o} (Stratonovich) calculus form of the limiting stochastic differential equation.  Similarly, It\^{o} calculus is invoked to interpret experiments wherein the time delay of the feedback is much larger than the noise correlation time.  

These results may also be reinterpreted within the framework of our shot noise formalism.  When the inertial or feedback time scales are much shorter than the noise decorrelation time, one can ignore the former and focus on the latter using the logic that lead us to 
the Stratonovich calculus interpretation.  However, when the inertial or feedback time scales are much longer than the noise decorrelation time, the duration $w$ of the shot noise is no longer equivalent to the decorrelation time scale of the noise. Rather, in order for the principle dynamics to be valid, we must interpret $w$ as the inertial or feedback time scale.  Hence, during a time increment $w$, we must assume that two consecutive random events $n_1$ and $n_2$ are independent, $P(n_1, n_2)=P(n_1)P(n_2)$, which is analogous to the equation (\ref{eqn:itotime}). Therefore, despite the colored noise in this case, It\^{o} calculus prevails. Furthermore, this description provides an unambiguous interpretation of
the discrete nature of a system \cite{horsthemke2006}, which is believed to be a central criterion for the use of It\^{o} calculus.      

\section{Conclusion}

When a system is described as having multiplicative noise, the criterion for choosing which stochastic calculus is most appropriate--It\^{o} or Stratonovich--has been been a source of confusion and great discussion. In most areas of physical science, where white noise is defined in terms of a $\delta$-function autocorrelation, 
Stratonovich calculus is preferred, mainly due to the consistency with the results emerging from the Fokker-Planck equation and the fluctuation-dissipation theorem. In economics, where the Martingale property is considered as the most essential aspect of stochastic random variables, It\^{o} calculus is widely accepted and used in the development of models. On the other hand, in population dynamics, It\^{o} calculus is viewed as a proper continuous approximation of  an underlying discrete model in which the Martingale property is guaranteed by construction.  However, it would appear more prudent to choose the calculus depending on a clear set of objective considerations.

The core difference between the approaches is seen through the presence of the Martingale property in the multiplicative noise term of the stochastic Langevin equation, and hence manifests itself in the drift term of the associated Fokker-Planck equation. Thus, the principal ambiguity is associated with the ensemble mean of the product of two Brownian motion processes and in the original form it is difficult to assess the validity of the Martingale property. Here we used integral Taylor expansions to pinpoint the source of the deviation between the two calculi, which resides in the characteristics of the short time-scale noise process, the nature of which is described by the autocorrelation.  The approach allows one to see the origin of the Stratonovich calculus, which resides in the mid-point selection scheme that is synonymous with  including the effect of a finite autocorrelation.  

It appears that most realistic signals simulated by white noise have non-Markovian structure, that is, a {\em finite decorrelation
time} $\tau_c$.  When one writes down a model stochastic differential equation, one typically initially assumes that $\tau_c$ is very small relative to the characteristic
deterministic dynamical time scales, in which case white noise is a proper approximation. Thus, in such a case, the integral of the noise autocorrelation function and the variance of the noise are of a similar order, and the $\delta$-function autocorrelation is a good approximation that is consistent with Stratonovich's calculus. 
However, when the integral of the noise autocorrelation function is much smaller than the variance of the noise, the $\delta$-function autocorrelation is not an appropriate definition of white noise and It\^{o}'s calculus is appropriate.

Focusing on the noise itself, in the microscopic limit it is no longer Brownian but is instead represented by discontinuous finite-time signals, or shot noise.  A small but finite autocorrelation of shot noise can be characterized by a $\delta$-function, and the accumulation over time of its autocorrelation is represented by the mid-point selection procedure in the Riemann sum of Stratonovich calculus. This demonstrates that the origin of $\delta$-correlated noise and Stratonovich calculus is the {\em infinitesimal autocorrelation}
of the stochastic noise.  Thus, the consistency between the Fokker-Planck equation and Stratonovich calculus has the same physical source, the finite autocorrelation of the stochastic noise that is ignored on time scales long relative to the fluctuation time scales.  Therefore, a finite noise autocorrelation is the key criterion for choosing a stochastic calculus, but we must take care in interpreting its origin generally and on a case by case basis.  

%If the choice of a mathematical methodology is dependent upon a specific field like mathematics, physics or economics, it would be hard to get rid of  the ambiguity in the justification of the methodology. The distinction among various fields is quite unclear in most of cases because there are a lot of similar processes in different fields. The choice of a stochastic calculus should not be dependent upon the field, but the intrinsic characteristics of the process we are considering. The continuity suggested by Wong and Zakai \cite{wong1965} was not enough to clear out the ambiguity of the choice of a stochastic calculus.We hope that the physical interpretation of Stratonovich calculus in this research contributes to providing a compelling criterion to choose a proper calculus in a specific physical model.
\eject
\begin{acknowledgments}
W.M. thanks NASA for a graduate fellowship and J.S.W. thanks the John Simon Guggenheim
Foundation, the Swedish Research Council, and a Royal Society Wolfson Research Merit Award for
support.  
\end{acknowledgments}

%\bibliographystyle{unsrt}
%\bibliography{sto}
%\end{document}

\end{document}